# Some Issues In Securitization And Disintermediation.

Michael C. Nwogugu
Address: Enugu 400007, Enugu State, Nigeria.
Email: Mcn2225@aol.com; mcn2225@gmail.com.
Phone: 234 814 906 2100.

**Abstract[1].**
Securitization has become prevalent in many countries, and has substantial impact on government monetary policy and fiscal policy which have not yet been adequately analyzed in the existing literature. This article develops optimal conditions for efficient securitization, identifies constraints on securitization, and analyzes the interactions of capital-reserve requirements and securitization. This article introduces new decision models and theories of asset-securitization.

1. Existing Literature On Securitization.

Securitization (see: Nwogugu (2005a, 2005b); Schwarcz (2004)) has radically changed the structure of the housing, banking, real estate and other industries, and is assumed to reduce the cost of borrowing in various industries because it increases money available for lending. Most securitizations are "secondary-market securitizations", where banks and financial institutions buy existing assets and then securitize them – this is very different from "primary-market securitization" where borrowing 'new' capital (secured by assets not already owned) is the primary objective. Relevant issues are discussed in: DeMarzo (2005); Riddiough (2000); Ambrose (2000); Sing, Ong, Fan & Sirmans (2004); Krainer (2002); Nofhaft & Freund (2003); Naranjo & Toevs (2003); Passmore, Sparks & Ingpen (2002); Glaeser & Kallal (1997); Heuson, Passmore & Sparks (2001); Ghiradato & Marinacci (2001); Mansini & Pferschy (2004); Mansini & Speranza (2002); Schwarcz (2002); Schwarcz (2004); DeYoung & Rice (2004); Browning, Gronsberg & Liu (____); David (1997); DeMarzo (2005); Jobst (June 2003); Calomiris & Mason (2004); Shakespeare (2001); Sitkoff (2003); Thomas (June 2001); Iacobucci & Winter (2005);

---

[1] A later version of this article was published as:

Mahul (2001); Van Order R (___); Dam (Feb 2006); Kaiser & Tumma (2004); Van Order (___); Cox, Fairchild & Pedersen (2000); Winton (2001); Siegl & Tichy (2000); Doherty & Schlesinger (2002); Fan, Sing & Ong & Sirmans (2004); Cooper (2001); Carnahan (2005); Murray (2001); Kupiec & Nickerson (2004); Nothaft & Freund (2003); Batchvarov, Hani & Davies (2002); Kollar (1999); Dionne & Harchaoui (May 2003); Thomas (June 2001); Cook & Hendershott (1978); Bothwell & Merrill (Feb. 2005); Campbell & Hercowitz (Jan. 2006); Herring & Chatusripitak (2001); Turner (_____); Cowley & Cummins (2005); Dalton & Dziobek (2005); Ferguson (Sept. 2003); Ghosh, Nachane, Narain & Sahoo (2003); Haggard, Thompson & Colonna (2000); Han (1996); Herzog (2006); Illing & Paulin (August 2004); Miller (1998); Pennachi (2006); Piga (2000); Pollin (1991); Ramey & Shapiro (1997); Riddiough (1997); Rogerson (1987); Smith & Wang (2006); Van Der Heuvel (1999); Vymyatnina (2006); Fleisig (June 1996); Anderson, Gilbert & Powell (1989); Ketkar & Ratha (Nov. 2004). Also see: Roberts & Pashler (Sept. 2005); Granger, King & White (1995).

However, the existing literature on asset-securitization does not analyze the following issues:

**1)** The conditions under which securitization reduce banks' probabilities of bankruptcy.

**2)** Securitization within the context of monitoring costs.

**3)** Empirical studies of securitization omitted the following: a) incorporation of monitoring costs, **b)** the impact of sources of capital (bank deposits, Fed Funds, pension funds, etc.) on the decision to securitize**, c)** the impact of investor risk tolerance, perceptions of the economy on the choice between ABS/MBS and other investments, **d)** the conditions under which yields spreads from securitization will remain positive; **e)** the conditions under which banks are better of syndicating loans instead of securitizing them, **e)** distinctions between secondary-securitizations and primary-securitizations in the data used in empirical studies, **f)** distinctions between transaction time (time taken to originate, structure and sell ABS/MBS), across different securitization transactions, **g)** the industry/asset effect – the impact of securitization is very likely to vary across industries and assets**; h)** the performance of ABS/MBS over time is not adjusted for inflation and changes in interest rates.

**4)** The effect of knowledge and knowledge management on the securitization process within banks and institutional investors. Securitization is a specialized financing technique and only a relatively small number of

---

Nwogugu, M. (2007). Some Issues In Securitization And Disintermediation. *Applied Mathematics & Computation,* 186(2): 1031-1039.

financial institution employees can manage such transactions. Furthermore, although a significant portion of processes are automated, the quality still depends on the validity/quality of the computer programs.

**5)** The optimal volume of securitization for any specific bank.

**6)** Most of the empirical studies on securitization suffer from the data validation, methodological and theory testing problems explained in Roberts & Pashler (2005).

**7)** Another critical issue is the classification of securitizations as true-sales or as assignments (financing), which has significant tax and accounting ramifications. The criteria for classification as a true sale are stated in Financial Accounting Standards #140 (FAS 140, September 2000) – "Accounting for Transfers And Servicing Of Financial Assets And Extinguishment Of Liabilities". For classification as a true sale, FAS requires that the sponsor must not have any control over the collateral, and that the SPV be a qualifying SPV (QSPV). FASB Interpretation #46 (FIN 46, December 2003) states the criteria for qualification as an SPV (FIN 46 does not apply to SPVs). However, the existing literature contains somewhat very divergent views about the quality of classification criteria (criteria proposed both by FASB, SEC and private entities/persons), the economic impact of such classifications and the type of sanctions required to reduce propensity for fraud.

2. Models Of Securitization.

Although securitization as practiced in the US, is technically illegal under US law [see: Nwogugu (2005); Schwarcz (2004)], the volume of securitization transactions and associated journal articles continue to increase. This section develops models of securitization (under US laws) with the objective of determining the conditions under which securitization is beneficial or non-beneficial to banks. The type of securitization analyzed here is one in which the bank/sponsor pools its assets and then securitizes them. The principles and formulas can be applied to various types of loans, credit card pools, leases, intellectual property and other types of assets. Since Prepayment Rates and the volume of prepaid assets in ABS transactions are relatively small compared to total ABS dollar volume, one approach is to completely omit ABS Prepayment-Rates when determining the optimal conditions for securitization. In all true-sale securitizations, the sponsor/bank will have typically made most of its profits before Prepayment rates become an issue. In assignment securitizations, the sponsor-bank will also have made substantial profits before Prepayments become an issue. Prepayments and changes in Prepayment rates have not been negative considerations to ABS investors, who typically understand prepayment mechanics and how to hedge them. Hence, for most financial institutions, the decision about whether or not to securitize is almost independent of ABS Prepayment Rates. Mansini & Speranza (2002); Mansini & Pferschy (2004); Murray (2001); Passmore & Sparks (2000); Winton (2001). On dynamical systems, see: Beer (2000); Dellnitz & Junge (1999); Moore (1991); Friedman & Sandler (1996); Iacus (2001); Van Gelder (1998); Naranjo & Toevs (2003).

Let:
$I_b$ = interest rate paid by banks to obtain deposits
$I_i$ = average interest rate paid by borrowers on mortgage loans
$I_f$ = banks' cost of borrowing funds from the Federal Reserve (Central Bank)
$I_s$ = PV of Interest paid on securitized debt, expressed as a percentage of the debt
$I_{spv}$ = Interest rate for securitized debt
$I_{lc}$ = PV of a bank's average transaction cost to make a loan – includes overhead, fixed costs per transaction and variable costs per transaction; expressed as a percentage of the loan amount.
$I_m$ = expected monitoring costs to monitor a loan that was not securitized; expressed as a percentage of the loan amount.
$I_{ms}$ = expected monitoring costs to monitor a loan that was securitized; expressed as a percentage of the loan amount.
$I_{ts}$ = transaction costs to securitize a loan; amortized over loan term and then expressed as a percentage of the loan amount.
$I_g$ = a bank's spread on a loan without securitization – including operating expenses
$I_{gs}$ = a bank's spread on a loan with securitization (including monitoring costs and operating expenses).
$I_l$ = average expected percentage credit loss per un-securitized loan (expressed as a percentage of the loan amount).
$I_{ls}$ = average expected percentage credit loss per securitized loan (expressed as a percentage of the loan amount).
$I_r$ = average residual rate in securitized transactions = $I_r = (I_i - I_s)$
$I_{rc}$ = average expected percentage credit loss on securitized debt
$I_{ci}$ = average cost of credit insurance in securitization; (expressed as a percentage of the loan amount).
$I_{cs}$ = average profits earned from syndicating unsecuritized loan; amortized over loan term and expressed as percentage of loan principal.
$I_{dr}$ = average periodic post default cost of loan resolution; (expressed as a percentage of the loan amount).
$P$ = loan amount.
$S$ = percentage of loan securitized. S = 1 in true-sales securitizations. $S \in (0,1)$
$S_s$ = percentage of original loan realized as new loan in securitization. $S_s \in S$.
$S_i$ = gain or loss on sale expressed as a percentage of the loan balance (applies only if transaction is deemed a true sale).
$T_{ts}$ = bank's tax rate in true-sale securitization
$T_a$ = bank's tax rate in assignment securitization
$T_{ls}$ = bank's tax rate in loan syndication
$T_o$ = bank's ordinary tax rate
$\psi_{bbs}$ = bank's probability of bankruptcy before securitization transaction (Some borrowers may have a higher credit rating than the bank).
$\psi_{bsa}$ = bank's probability of bankruptcy after securitization transaction.
$\psi_{bbl}$ = bank's probability of bankruptcy before loan syndication transaction. (Some borrowers may have a higher credit rating than the bank).
$\psi_{bal}$ = bank's probability of bankruptcy after loan syndication transaction
$\psi_s$ = bank's probability of successful loan syndications (syndication of pool of loans).

$V_s$ = Volume of total securitized mortgage debt in the economy
$V_d$ = Volume of total bank deposits in the economy
$V_f$ = volume of debt borrowed by all banks in the economy from the Central Bank (US Fed)

$E_1, E_2, E_3, E_4$ = with respect to organizational form of SPV, there are various states that correspond to the various possible corporate forms of the SPV. The SPV can be an LLC ($E_1$), LLP ($E_2$), C-corporation ($E_3$) or trust ($E_4$), each of which has its unique tax benefits/problems, and degree of asset protection and limited liability under state laws. $E_0$ is the optimal state during the life of the SPV, such that deviations from $E_o$ ($|E_0-E_i|$) measures the loss and opportunity cost arising from use of sub-optimal corporate structures as SPVs – such loss is typically in the form of increased legal fees, increased losses due to low limited-liability, and the value of extra time spent in dispute resolution.

$F_f, F_s$ = with respect to the form of financing, there are two states of the world – $F_f$, where the transaction is an assignment/financing, and $F_s$, where the transaction is a true sale. In an assignment, the sponsor does not subsidize the SPV, but only substitutes impaired collateral. Where $F_f$ is the best/ideal state but $F_s$ is being used, the transaction incurs losses, and vice versa; but where the correct transaction format is used, there is no loss/gain. Hence the value of the transaction format choice is $\mathbf{F} = \text{Max}[|F_f - F_s|, 0]$.

$A$ = value of the adverse selection problem. Where the transaction is an assignment and the sponsor is the servicer, the sponsor will typically substitute impaired collateral, and hence has an incentive not to profer the best collateral as replacement. Where $C_0$ = impaired collateral, $C_i$.....$C_n$ = value of replacement collateral, and $C_a$ is the value of average-quality collateral available for replacement and its assumed that collateral consist of units of the same or similar sizes; $P_i$ = probability of there being impaired collateral; $P_a$ = probability that the servicer/sponsor will offer medium-quality or low-quality collateral as replacement for the impaired collateral. $\mathbf{A} = {_0\int^t} \{\Sigma(C_a)*P_i (1- P_a)\}$

$t$ = the period of time that it takes the bank to do one securitization transaction.

Without securitization, the bank's profit in time $t$ =

$$I_g = [I_i - I_{lc} - I_l - I_m - \text{Max}(I_f, I_b)] + \Sigma I_{gf}$$

Where the last term ($\Sigma I_{gf}$) represents the present value of projected profits from additional loans based on this loan and funded by borrowing from the US Fed, using this loan as collateral.

With securitization, the bank's profit in time $t$ is equal to:

$$I_{gs} = [\{(I_r-I_c-I_{ci}-I_{ms}-I_{ts})(P)(S)(S_s)\} + \{(1-S_s)(S)(I_i-I_{lc}-I_{ts})\} + \{(1-S)(I_i - I_{lc} - I_l - \text{Max}(I_f,I_b))\} + \Sigma I_{gs}] - \text{Max}[|F_f - F_s|, 0] - \text{Max}[0, (|E_0-E_1|), (|E_0-E_2|), (|E_0-E_3|), (|E_0-E_4|)] - A$$

Where $\Sigma I_{gsf}$ represents the present value of future gains from subsequent loans and securitizations after time $t$. Oldfield (2000). Passmore & Sparks (2000).

Hence for the bank to profitably enter into securitization for time $t$, then the following conditions must exist:
$(1-\text{Max}\{T_{ts},T_a\})$
**1.** $I_i > I_{spv} > \text{Max}(I_b, I_f, 0)$
**2.** $[\{(I_r-I_c-I_{ts}-I_{ci}-I_{ms})(1-\text{Max}\{T_{ts},T_a\})(P)(S)(S_s)\} + \{(1-S_s)(1-\text{Max}\{T_{ts},T_a\})(S)(I_i-I_{lc}-I_{ts})\} + \{(1-S)(1-\text{Max}\{T_{ts},T_a\})(I_i - I_{lc} - I_l - \text{Max}(I_f,I_b))\} + S_i + \Sigma I_{gs}] > [\{I_i - I_{lc} - I_{ls} - \text{Max}(I_f,I_b) - I_m\}(1-\text{Max}\{T_{ls},T_o\})*P]$
**3.** $[\{(I_r-I_c-I_{ts}-I_{ci}-I_{ms})(P)(S)(S_s)\} + \{(1-S_s)(S)(I_i-I_{lc}-I_{ts})\} + \{(1-S)(I_i - I_{lc} - I_l - \text{Max}(I_f,I_b))\} + S_i + \Sigma I_{gs}] > [\{I_i - I_{lc} - I_{ls} - \text{Max}(I_f,I_b) - I_m\}*P]$
**4.** $(\partial I_{gs}/\partial I_{ts}) > \text{Max}[(\partial I_g/\partial I_i), 1]$

**5.** $(\partial P_{rp}/\partial I_g) < 1$

**6.** $\text{Max}[(\partial I_g/\partial I_m), 1] < (\partial I_{gs}/\partial I_{ms})$

**7.** $I_{rc} > I_i$ ; $I_{gs} > 0$; $\text{Max}[I_f, I_{spv}, I_b] < I_i$;

**8.** $\partial^3 A/\partial F^3 > 1$

**9.** $\partial^2 I_{gs}/\partial I_f^2 > \text{Max}[(\partial I_{gs}/\partial I_f), (\partial^2 I_{gs}/\partial I_b^2), 1]$; and $\partial I_{gs}/\partial I_f > \text{Max}[\partial I_{gs}/\partial I_b, 1]$

**10.** $\partial^3 I_b/\partial P_{rp}^3 < \text{Min}[(\partial^3 I_i/\partial P_{rp}^3), 0]$

**11.** $P_{rp} > P_{ra}$; $(P_{rp} | \text{Max}(E1, E2, E3, E4) > P_{ra} | \text{Max}(E1, E2, E3, E4)$

**12.** $I_m > I_{ms}$; $\{(I_m | \text{Max}(E1, E2, E3, E4) > I_{ms} | \text{Max}(E1, E2, E3, E4)$

**13.** $I_m | \text{Max}(F_f, F_s) > I_{ms} | \text{Max}(F_f, F_s)$;

**14.** $I_m | \text{Max}(| F_f - F_s |, 0). > I_{ms} | \text{Max}(| F_f - F_s |, 0)$; and $P_{rp} | \text{Max}(| F_f - F_s |, 0) > P_{ra} | \text{Max}(| F_f - F_s |, 0)$;

**15.** $\partial I_l/\partial t_m > \text{Max}[\partial I_{ls}/\partial t_m, 0]$

**16.** $[\{(I_r - I_c - I_{ts})(P)(S)(S_s)\} + \{(1 - S_s)(S)(I_i - I_{lc} - I_{ts})\} + \{(1-S)(I_i - I_{lc} - I_l - \text{Max}(I_f, I_b))\} + \sum I_{gs}] < 0$

**17.** $[\{(I_r - I_c - I_{ts})(P)(S)(S_s)\} + \{(1 - S_s)(S)(I_i - I_{lc} - I_{ts})\} + \{(1-S)(I_i - I_{lc} - I_l - \text{Max}(I_f, I_b))\}] < 0$

**18.** $\{I_i - I_s - I_{ts} - I_c\} < 0$

**19.** $(\partial^2 V_s/\partial V_d \partial V_f) > 1$

**20.** $\partial I_{spv}/\partial V_s > \text{Max}[\partial I_b/\partial V_d, 0]$;

**21.** $\partial^2 I_{spv}/\partial V_d \partial V_f > 0$; $\partial^2 I_b/\partial V_d \partial V_f > 0$;

**22.** $\partial^3 I_{gs}/\partial V_s \partial V_d \partial V_f > 0$ ;

**23.** $\partial I_{gs}/\partial V_s > 1$; $\partial I_{gs}/\partial V_f \partial V_d > 0$

**24.** $\partial I_f/\partial V_f > \partial I_d/\partial V_d$

**25.** $Ig_s > I_{gs}$; $(\partial I_g/\partial I_{cs}) > \text{Min}[(\partial I_{gs}/\partial I_{ls}), 1]$; $(\partial^2 I_{gs}/\partial I_{cs}^2) > \text{Min}[(\partial^2 I_{gs}/\partial^2 I_{ls}^2), 1]$

**26.** $(\partial S/\partial I_i) > \text{Max}[(\partial S/\partial I_{spv}), 1]$; $(\partial^2 S/\partial I_i^2) > \text{Max}[(\partial^2 S/\partial I_{spv}^2), 1]$;

**27.** $\partial I_{ms}/\partial It_{ts} > 1$

**28.** $\partial S/\partial(\psi_{bbs} - \psi_{bsa}) > \text{Max}[\partial P/\partial(\psi_{bbl} - \psi_{bal}), 1]$

**29.** $\partial^2 S/\partial S_s \partial(\psi_{bbs} - \psi_{bsa}) > \text{Max}[\partial^2 S/\partial(\psi_{bbs} - \psi_{bbs})^2, 0]$

In order to create a decision model, weights $(x_1 \ldots x_n)$ and scores $(y_1 \ldots y_n)$ can be assigned to each of these conditions, subject to $\sum x_n = 1$; and the sum of the weighted scores can serve as the decision indicator. The decision maker can then choose/apply various thresholds between 0 and 1.

3. The Bank's Objective Function.
    The bank's multi-criteria objective function will be:
Max $[\{(I_r - I_c - I_{ts} - I_{ci} - I_{ms})(P)(S)(S_s)\} + \{(1 - S_s)(S)(I_i - I_{lc} - I_{ts})\} + \{(1-S)(I_i - I_{lc} - I_l - \text{Max}(I_f, I_b))\} + \sum I_{gs}]$

Hence, banks and financial institutions will continue to engage in securitization until their transaction costs, credit loss risks and spreads preclude further transactions. Carlstrom & Samolyk (1993); Donahoo & Shaffer (1991); Hendershott & Van Order (1989); Greenbaum & Thakor (1987). Generally, as the volume of securitized mortgage debt ($V_s$) increases, its expected that:

$I_{spv} \to I_b$, and
$\partial I_s/\partial I_b < 0$; until investors will become indifferent between $I_{spv}$ and $I_b$.

Dionne & Harchaoui (2003) studied the impact of securitization and credit risk on banks' capital and found that : 1) securitization has negative effects on both Tier-1 capital ratios and Total risk-based capital ratios; 2) there is a positive statistically significant relationship between securitization and banks' risk. Similarly, some researchers found that under the current capital requirements for credit risk, banks may be induced to (and have substantial incentives to) shift to more risky assets (loans). Donahoo & Shaffer (1991).

Its conjectured here that securitization negates bank capital-reserve-requirements regulations by providing incentives for banks to take excessive risks. So long as banks can purchase and securitize highly risky assets through true-sale securitizations, current bank capital-reserve-requirements in most jurisdictions (even those that purport to incorporate the risk of banks' off-balance sheet activities) don't preclude financial institution failures. Greenbaum & Thakor (1987); Donahoo & Shaffer (1991); Murray (2001); Winton (2001); Naranjo & Toevs (2003).

4. Optimal Amount Of Securitization For A Bank

As explained in the preceding sections, the decision to securitize assets (both true-sale and assignment securitizations) is not affected by existing bank capital-reserve requirements. Berger & Udell (1994); Chakraborty & Ray (2006); Allen & Carletti (2006); Cooper (2001); Browning, Gronsberg & Liu (_____); DeYoung & Rice (2004); Krainer (2002); Naranjo & Toevs (2003); Greenbaum & Thakor (1987); Donahoo & Shaffer (1991); Cosimano & McDonald (1998); Naranjo & Toevs (2003). Its assumed that all securitization transactions earn net-spreads (net of monitoring costs, transaction costs, cost of credit losses, cost of credit enhancements, and administration costs) for the bank-sponsors, otherwise the bank-sponsor will not undertake the transactions. The months and months and months that the In most securitizations, credit losses are typically low (less than 5%), and standard over-collateralization and true-sales eliminates any recourse to the sponsor bank. Thus, the magnitude and probability of Asset Substitution are not constraints on, or even minor causal factors for the bank's optimal securitization volume. Theoretically, a bank's positive net-spreads from securitization are not perpetual but is limited/constrained by the following factors:

**a)** The available volume of assets/loans that can be securitized is finite;

**b)** Investors are knowledgeable and can buy other assets instead of ABS/MBS and or re-allocate capital from ABS-MBS to other securities/assets;

**c)** There are other banks and financial institutions and sponsors that will issue ABS/MBS and seek similar spreads.

Therefore, the optimal amount of securitization for a specific bank is determined mostly by external factors that affect the overall banking/capital markets; and is a function of the following:

**a)** The available volume of assets/loans that can be securitized, which is finite;

**b)** Investors are knowledgeable and can buy other assets instead of ABS/MBS;

**c)** There are other banks and financial institutions and sponsors that will issue ABS/MBS;

**d)** The asset class (ie. – leases vs. mortgages vs. healthcare receivables, etc.). The optimum securitization volume for any varies by assets class, because class characteristics vary substantially, and relative risk is evaluated differently for each asset class.

**e)** The investors' average propensity to substitute a specific class of ABS/MBS with another security;

**f)** The marginal yield on ABS/MBS;

g) the ratio of the bank's total capital to the sum of the total capital of all banks/sponsors/financial institutions that can issue ABS/MBS in that asset class; and a former

**h)** The bank's long run market share in the ABS/MBS sector (this is a function of the bank's marketing prowess, knowledge about origination/structuring/hedging/trading of ABS/MBS, and social capital).

Hence the optimal ABS/MBS volume in an asset-class for a bank, occurs at that ABS/MBS dollar-volume where the following conditions exist:

**a)** $M_{mys} = M_{cm}$; $S_c \geq S_h$

**b)** $[(V_m)*(S_c)] \geq [(S_h)*(V_p)]$

**c)** $V_p > (S_c * V_m)$

**d)** $M_{mys} \geq [(M_{cm}*R_{sn}*\Phi) + (M_{ays}*R_{ss}*\Omega)]$

**e)** $(\partial R_{ss}/\partial V_m) < 1$; and $(\partial^2 R_{sn}/\partial V_m^2) < 0$;

**f)** $\partial R_{sn}/\partial V_p < 0$; and $(\partial^2 R_{ss}/\partial V_m^2) < 0$;

**g)** $\mathbf{M_{mys}} \geq (M_{cm} * R_{ss})$

Where:

**Φ, Ω** = Φ is the relative impact of the non-ABS/MBS capital markets; and Ω is the relative impact of the ABS/MBS market. $\Phi + \Omega = 1$; $0 < \Phi < 1$; $0 < \Omega < 1$.

**$M_{mys}$** = the marginal Modified-Duration on the average ABS/MBS in the asset class (ie. leases vs. business loans vs. mortgages vs. healthcare receivables vs. aircraft receivables, etc.). $-\infty < M_{mys} < +\infty$.

**$M_{cm}$** -= the average Modified-Duration of comparable non-ABS/MBS securities with similar characteristics (duration, coupon, term, etc.). $-\infty < M_{cm} < +\infty$.

**$M_{ays}$** = the average Modified-Duration of comparable ABS/MBS securities in the same asset class, with similar characteristics (duration, coupon, term, etc.). $-\infty < M_{ays} < +\infty$.

**$R_{sn}$** = the average investor's marginal rate of substitution of ABS/MBS for another non-ABS/MBS security. $-\infty < R_{sn} < +\infty$.

**$R_{ss}$** = the average investor's marginal rate of substitution of ABS/MBS for another ABS/MBS in the same asset class. $-\infty < R_{ss} < +\infty$.

**$V_m$** = actual total dollar volume of ABS/MBS market segment. This includes only securitized assets in the segment. $0 < V_m < +\infty$.

**$V_p$** = potential dollar volume of ABS/MBS market segment – this includes both securitized and un-securitized assets in the asset-class/segment. $0 < V_p < +\infty$. In reality, this amount varies continuously, but for this analysis, its assumed that the optimal amount of securitization is as of a specific point in time.

**$S_h$** = the bank's historical market share in the ABS/MBS segment. $0 < S_h < 1$.

**$S_c$** = the bank's current market share in the ABS/MBS segment. $0 < S_c < 1$.

5. Conclusion.
Although Securitization is illegal (see Nwogugu (2005)), it has radically altered the structures of the financial services industry and the real estate industry, and coupled with the Internet, can have a major impact on the characteristics/regulation of property acquisition/disposition processes. This article has shown that securitization may or may not be beneficial for banks, depending on interest rates, transaction costs, monitoring costs, information-diffusion rates, marginal propensity to substitute ABS/MBS for other assets, market completeness, and the rate at which arbitrage opportunities dissipate in the capital markets.

6. Bibliography.
**1.** Allen F & Carletti E (2006). Credit Risk Transfer And Contagion. *Journal Of Monetary Economics*, 53: 89-111.
**2.**. Ambrose B (2000). Local Economic Risk Factors And The Primary And Secondary Mortgage Markets. *Regional Science & Urban Economics*, 30(6): 683-701.
**3.** Anderson R, Gilbert C & Powell A (1989). Securitization And Commodity Contingency In International Lending. *American Journal Of Agricultural Economics*, 71(2): 523-530.
**4.** Batchvarov A, Hani C & Davies W (2002). Mortgage Securitization, Default Swaps And Financial Guarantees; Who Buys, Who Sells? *Housing Finance International*, 17(1):43-51.
**5.** Beer R (2000). Dynamic Approaches To Cognitive Science. *Trends In Cognitive Sciences*, 4(3):91-94.
**6.** Berger A & Udell G (1994). Did Risk-Based Capital Allocate Bank Credit And Cause A "Credit Crunch" In The United States ?. *Journal of Money Credit & banking*, 26:
**7**. Bothwell J & Merrill S (Feb. 2005). Regulation And Supervision Of Mortgage Lending In The Emerging Markets: An Assessment In Bulgaria And Romania. Working paper, The Urban Institute, Washington DC, USA.
**8.** Browning E, Gronsberg T & Liu L (____). Alternative Measures Of The Marginal Cost Of Funds. *Economic Inquiry*, 38(4): 591-599.
**9.** Calomiris C & Mason J (2004). Credit Card Securitization And Regulatory Arbitrage. *Journal OF Financial Serices Research*, 26:5-28.
**10.** Campbell J & Hercowitz Z (Jan. 2006). *The Role OF Collateralized Household Debt In Macroeconomic Stabilization*. Working Paper.
**11.** Carlstrom C & Samolyk K A (1993). Examining The Micro-Foundations Of Market Incentives For Asset-Backed Lending. *Economic Review – Federal Reserve Bank of Cleveland*, 29(1):27-39.
**12.** Carnahan S (2005). Home Equity Line of Credit Securitization: Issuer Issues. *Journal of Structured Finance*, 11(3):30-32.
**13.** Chakraborty S & Ray T (2006). Bank-Based versus Market based Financial Systems: A Growth-Theoretic Analysis. *Journal Of Monetary Economics*, 53: 329-350.
**14**. Cook T & Hendershott P (1978).. The Impact Of Taxes, Risk And Relative Security Supplies On Interest Rate Differentials. *Journal Of Finance*, 33:1173-1186.
**15.** Cooper I (2001). Comments On The Paper By Ernst Von Thaden: "Liquidity Creation Through Banks And Markets: A Theoretical Perspective On Securitization". *Economic Notes*, 29(3): 393-397.
**16.** Cosimano T & McDonald W (1998). Whats Different Among Banks ? *Journal Of Monetary Economics*, 41:57-70.
**17.** Cowley A & Cummins D (2005). Securitization Of Life Insurance Assets And Liabilities. *Journal of Risk and Insurance*, 72(2):193-226.
**18.** Cox S, Fairchild J & Pedersen H (2000). Economic Aspects Of Securitization Of Risk. *ASTIN Bulletin,* 30(1): 157-193.
**19.** Dalton J & Dziobek C (2005). Central Bank Losses And Experiences In Selected Countries. IMF Working Paper WP/05/72.
**20.** Dam K (Feb 2006). Credit Markets, Creditor's Rights And Financial Development. University Of Chicago Law School, Working Paper.
**21.** David A (1997). Controlling Information Premia By Repackaging Asset Backed Securities. *Journal Of Risk & Insurance*, 64(4):619-648.

**22.** Dellnitz M & Junge O (1999). On The Approximation Of Complicated Dynamical Behavior. *SIAM Journal Of Numerical Analysis*, 36(2):491-515.
**23.** DeMarzo P (2005). The Pooling And Tranching Of Securities: A Model of Informed Intermediation. *Review Of Financial Studies*, 18(1): 1-35.
**24.** DeYoung R & Rice T (2004). How Do Banks Make Money ? A Variety Of Business Strategies. *Federal Reserve Bank Of Chicago – Economic Perspectives*, pp. 52-68.
**25.** Dionne G & Harchaoui T (May 2003). Bank's Capital, Securitization And Credit Risk: An Empirical Evidence For Canada. Working paper # CREF 03-01, Les Cahiers Du CREF, Canada.
**26.** Doherty N & Schlesinger H (2002). Insurance Contracts And Securitization. *Journal Of Risk And Insurance*, 69(1): 45-62.
**27.** Donahoo K & Shaffer S (1991). Capital Requirements And the Securitization Decision. *The Quarterly review of Economics & Business*, 31(4):12-24.
**28.** Fan G, Sing T & Ong S & Sirmans C (2004). Governance And Optimal Financing For Asset Backed Securitization. *Journal of Property Investment & Finance*, 22(5); 414-434.
**29**. Ferguson R (Sept. 2003). Capital Standards For Banks: The Evolving Accord. *Federal Reserve Bulletin*.
**30.** Fleisig H (June 1996). Secured Transactions: The Power of Collateral. *Finance & Development*, pp. 44-49.
**31.** Friedman Y & Sandler U (1996). Evolution Of Systems Under Fuzzy Dynamic Laws. *Fuzzy Sets And Systems*, 84:61-72.
**32.** Ghiradato P & Marinacci M (2001). Risk Ambiguity And The Separation of Utility And Beliefs. *Mathematics Of Operations Research,* 26(4):864-890.
**33.** Ghosh S, Nachane D, Narain A & Sahoo S (2003). Capital Requirements And Bank Behavior: An Empirical Analysis Of Indian Public Sector Banks. *Journal Of International Development*, 15:145-156.
**34.** Glaeser E & Kallal H (1997). Thin Markets, Asymmetric Information And Mortgage Backed Securities. *Journal Of Financial Intermediation*, 6(1): 64-81.
**35.** Granger C, King M & White H (1995). Comments On testing Economic Theories And The Use Of Model Selection Criteria. *Journal Of Econometrics*, 67:173-187.
**36.** Greenbaum S & Thakor A (1987). Bank Funding Modes: Securitization Versus Deposits. *Journal of Banking & Finance*, 11(3):379-399.
**37.** Haggard M, Thompson M & Colonna S (2000). Exploring The Capital Markets And Securitization For Renewable Energy Projects. Working Paper, UK.
**38.** Han J (1996). To Securitize Or Not To Securitize ? The Future Of Commercial Real Estate Debt Markets. *Real Estate Finance*, Summer 1996, pp. 71-80.
**39** . Hendershott P & Van Order R (1989). Integration of Mortgage and Capital Markets and the Accumulation of Residential Capital. *Regional Science and Urban Economics*, 19(2):189-209.
**40.** Herring R & Chatusripitak N (2001). *The Case Of The Missing Market: The Bond Market And Why It Matters For Financial Development*. Wharton Financial Institutions Center. Working Paper.
**41.** Herzog B (2006). Coordination Of Fiscal And Monetary Policy In CIS Countries: A Theory Of Optimum Fiscal Area ? *Research In International Business & Finance*, 20(2): 127-274.
**42.** Heuson A, Passmore W & Sparks R (2001). Credit Scoring And Mortgage Securitization: Implications For Mortgage Rates And Credit Availability. *Journal Of Real Estate Finance And Economics*, 23(3): 337-363.
**43.** Iacobucci E & Winter R (2005). Asset Securitization And Assymetric Information., *Journal Of Legal Studies*, 34:161-206.
**44.** Iacus S (2001). Efficient Estimation Of Dynamical Systems. *Nonlinear Dynamics And Econometrics*, 4(4):213-226.
**45.** Illing M & Paulin G (August 2004). *The New Basel Capital Accord And The Cyclical Behavior Of Bank Capital*. Bank of Canada, Working Paper #2004-30.
**46.** Jobst A (June 2003). *Risk Sharing In Securitization – The Function Of The First Loss Provision As A Signaling Device*. Working paper, June 2003.
**47.** Thomas H (June 2001). Effects Of Asset Securitization On Seller Claimants. Working Paper Chinese University of Hong Kong.
**48.** Kaiser M & Tumma S (2004). Take Or Pay Contract Valuation Under Price And Private Uncertainty. *Applied Mathematical Modelling*, 28: 653-676.

**49.** Ketkar S & Ratha D (Nov. 2004). *Development Financing During Crises: Securitization Of Future Recievzbles*. World Bank, Research Working Paper #S2582.
**50.** Krainer R (2002). Banking In A Theory Of The Business Cycle: A Model And Critique Of The Basle Accord On Risk-Based Capital Requirements For Banks. *International Review of Law and Economics*, 21(4):413-433.
**51.** Kupiec P & Nickerson D (2004). Assessing Systemic Risk Exposure from Banks and GSEs Under Alternative Approaches to Capital Regulation. *Journal of Real Estate Finance and Economics*, 28(2-3):123-145.
**52.** Mahul O (2001). Managing Catastrophic Risk Through Insurance And Securitization. *American Journal Of Agricultural Economics*, 83(3): 656-661.
**53.** Mansini R & Speranza M G (2002). A Multidimensional Knapsack Model For Asset-backed Securitization. The *Journal of the Operational Research Society*, 53(8): 822-832.
**54.** Mansini R & Pferschy U (2004). Securitization Of Financial Assets: Approximation In Theory And Practice. *Computational Optimization and Applications*, 29(2):147-171.
**55.** Miller G P (1998). An Interest Group Theory Of Central Bank Independence. *Journal Of Legal Studies*,
**56.** Moore C (1991). Generalized Shifts: Unpredictability And Un-Decidability In Dynamical Systems. *Nonlinearity*, 4:199-230.
**57.** Murray A (2001). Has Securitization Increased Risk To The Financial System? *Business Economics*, 36(1):63-67.
**58.** Naranjo A & Toevs A (2003). The Effects Of Purchase Of Mortgages And Securitization By Government Sponsored Agencies On Mortgage Yield Spreads And Volatility. *Journal Of Real Estate Finance & Economics*, 25(2/3): 173-195.
**59.** Nofhaft F & Freund J (2003). The Evolution Of Securitization In Multifamily Mortgage Markets And Its Effect On Lending Rates. *Journal Of Real Estate Research*, 25(2): 91-101.
**60.** Nwogugu M (2005a). Securitization Is Illegal. Working Paper Series, www.ssrn.com.
**61.** Nwogugu M (2005b). Some New Theories Of Securitization. Working Paper Series, www.ssrn.com.
**62.** Oldfield G (2000). Making Markets For Structured Mortgage Derivatives. *Journal of Financial Economics*, 57(3):445-471.
**63.** Passmore W & Sparks R & Ingpen J (2002). GSEs, Mortage Rates And The Long Run Effects Of Mortgage Securitization. *Journal Of Real Estate Finance & Economics*, 25(2/3): 215-242.
**64.** Passmore W & Sparks R (2000). Automated Underwriting And The Profitability Of Mortgage Securitization. *Real Estate Economics*, 28(2):285-305.
**65.** Pennachi G (2006). Deposit Insurance, Bank Regulation And Financial System Risks. *Journal of Monetary Economics*, 53:1-30.
**66.** Piga G (2000). Dependent And Accountable: Evidence From The Modern Theory Of Central Banking. *Journal of Economic Surveys*, .
**67.** Pollin R. (1991). Two Theories Of Money Supply Endogeneity: Some Empirical Evidence. *Journal of Post-Keynesian Economics*, 13(3):366-395.
**68.** Ramey V & Shapiro M (1997). *Costly Capital Reallocation And The Effects of Government Spending*. Carnegie-Rochester Conference Series On Public Policy, 48:145-194.
**69.** Riddiough T (1997). Optimal Design And Governance Of Asset-Backed Securities. *Journal Of Financial Intermediation*, 6:121-152.
**70.** Riddiough T (2000). Forces Changing Real Estate For At Least A Little While: Market Structure And Growth Prospects Of The Conduit-CMBS Market. *Real Estate Finance*, 17: 52-61.
**71.** Roberts S & Pashler H (2000). *How persuasive Is A Good Fit ? A Comment On Theory Testing*. University Of California, Department of California, Paper 763.
**72.** Rogerson B (1987). An Equilibrium Model Of Sectoral Reallocation.. *Journal Of Political Economy*, 95:824-834.
**73.** Schwarcz S (2002). The Universal Language Of International Securitization. *Duke Journal Of Comparative And International Law*, 12:285-300.
**74.** Schwarcz S (2004). Is Securitization Legitimate ? *International Financial Law Review*. 2004 Giude To Structured Finance, pp. 115.
**75.** Siegl T & Tichy R (2000). Ruin Theory With Risk Proportional To The Free Reserve And Securitization. *Insurance, Mathematics & Economics*, 26(1): 59-73.

**76.** Sing T, Ong S & Sirmans C (2004). Analysis Of The Credit Risks In Asset Backed Securitization Transactions In Singapore. *Journal Of Real Estate Finance & Economics*, 28(2/3): 235-253.
**77.** Sitkoff R (2003). *Trust Law, Corporate Law And Capital Market Efficiency*. Working Paper #20, John Olin Center For Law And Economics, University Of Michigan.
**78.** Shakespeare C (2001). *Do Managers Use Securitization Volume And Fair Value Estimates To Hit Earnings Targets* ? Working Paper, University Of Michigan.
**79.** Smith A & Wang C (2006). Dynamic Credit Relationships In General Equilibrium. *Journal Of Monetary Economics*, 53:847-877.
**80.** Thomas H (June 2001). *Effects Of Asset Securitization On Seller Claimants*. Working Paper, McMaster University, and Chinese University of Hong Kong.
**81.** Turner P (_____). Bond Markets In Emerging Economies: An Overview Of Policy Issues. BIS Papers, #11.
**82.** Van Der Heuvel S (1999). Does Bank Capital Matter For The Transmission Of Monetary Policy ? *Economic Policy Review, Federal Reserve Bank Of New York*, 8(1).
**83.** Van Gelder T (1998). The Dynamical Hypothesis In Cognitive Science. *Behavioral And Brain Sciences*, 21:615-665.
**84.** Van Order R (___). *Securitization And Community Lending: A Framework And Some Lessons From The Experience In The Us Mortgage Market*. Federal Reserve Bank Of San Francisco.
**85**. Vymyatnina Y (2006). How Much Control Does Bank Of Russia Have Over Money Supply ? *Research In International Business & Finance*, 20(2): 131-144.
**86.** Winton A (2001). Institutional Liquidity Need And The Structure Of Monitored Finance. *Review Of Financial Studies*, 16:1273-1313.